\documentclass[11pt,a4paper]{article}
\usepackage[hyperref]{emnlp-ijcnlp-2019}
\usepackage{times}
\usepackage{latexsym}
\usepackage{amsmath,amssymb,amsfonts}
\usepackage{algorithmic}
\usepackage{graphicx}
\usepackage{textcomp}
\usepackage{xcolor}
\usepackage{times}
\usepackage{latexsym}
\usepackage{algorithm2e}
\usepackage{booktabs} 
\usepackage{subfigure}
\usepackage{multirow}
\usepackage{url}
\usepackage{hyperref}
\usepackage{tabularx}
\SetKwComment{Comment}{$\triangleright$\ }{}

\SetCommentSty{mycommfont}

\usepackage{url}
\usepackage{url}

\aclfinalcopy 

\title{Using Semantic Role Knowledge for Relevance Ranking of Key Phrases in Documents: An Unsupervised Approach}

\author{Prateeti Mohapatra \\
  IBM Research \\  \And
  Neelamadhav Gantayat \\
  IBM Research \\
  {\tt \{pramoh01, neelamadhav, GaargiDasgupta\}@in.ibm.com} \\ \And
  Gargi B. Dasgupta \\ 
  IBM Research \\}

\date{}

\begin{document}
\maketitle
\begin{abstract}
  In this paper, we investigate the integration of
sentence position and semantic role of words
in a PageRank system to build a key phrase
ranking method. We present the evaluation results
of our approach on three scientific articles.
We show that semantic role information,
when integrated with a PageRank system, can
become a new lexical feature. Our approach
had an overall improvement on all the data sets
over the state-of-art baseline approaches.
\end{abstract}

\section{Introduction}
\label{intro}

Graph based approaches are currently the state-of-the-art for unsupervised key phrase ranking ~\cite{hasan-ng:2014}. 
TextRank ~\cite{Mihalcea04TextRank} and SingleRank ~\cite{Wan:2008} build a graph from the document and rank its nodes according to their importance. PositionRank uses a word position information approach to rank key phrases ~\cite{PosnRank:2017}. More recently, ~\cite{N18-2105} used a multipartite graph structure to represent documents as tightly connected sets of topic related candidates. For intra-topic key phrase ranking, they used features like position information of phrases, where, key phrases are promoted if they occur at the beginning of the document.

The above-cited papers make the assumption that words found early in a document are more important than words occurring later. However, this assumption is not always true. 
Consider the following example from a document: \\
\emph{Qualified planning of a power supply concept is the key to the efficiency of electric power supply.}
Here, word position would give higher relevance to {\it qualified planning}; however, intuitively  {\it electric power supply} should have got the higher relevance score for this sentence.



There can be other scenarios where sentences can be written or expressed in different ways for the same intention or meaning.  Take, for example, the following two sentences: \\
{\it Sent1: } Folic Acid can improve anxiety. \\
{\it Sent2: } Anxiety can be controlled by Folic Acid.


For {\it Sent2}, using word position, {\it Anxiety} gets a higher relevance than {\it Folic Acid}, since it appears earlier. However, in the first sentence {\it Folic Acid} gets a higher relevance than {\it Anxiety}, even though both sentences convey the same meaning.

Hence, lack of semantic information in the current system may yield incorrect results. In order to identify the relevance of a key phrase in a sentence, one needs to understand the event representations embedded in the text, and its lexical semantic relations within the sentence.
Motivated by this, we investigate here the effect of integrating sentence position of words and semantic role information for key phrase(s) relevance ranking. 



\section{Related Work}
\label{related}



{\it TextRank}~\cite{Mihalcea04TextRank}, {\it SingleRank}~\cite{Wan:2008} and {\it TopicRank}~\cite{Liu:2010} algorithms, to name a few, are the state-of-the-art techniques being used for graph-based key phrase extraction. 
Others have incorporated word position information as node weights to score key phrases ~\cite{PosnRank:2017,graph3}. 
Other approaches have also used word clustering and topic modeling techniques and then extracted key phrases from each topic ~\cite{graph2,N18-2105}.

None of the above approaches consider the semantic role of a word in a sentence, the heuristics being that the relevance of key phrases should be based on their corresponding roles in the sentence. 

\section{Semantic Role Labeling (SRL)}
\label{srl}
Semantic role labeling ~\cite{pradhan2004shallow, Gildea} is a shallow semantic parsing task describing who did what to whom, where, when, etc. For each predicate in a sentence, SRL identifies constituents which either play a semantic role (agent, patient, instrument, etc.) or act as an adjunct (location, manner, temporal etc.).  
Our SRL realization is based on the implementation by Roth ~\cite{roth2014composition}.


There are five semantic roles numbered from Arg0 to Arg5, and a secondary agent tag (ArgA) for proto-agents. ArgA is used with respect to a predicate where the verb indicates a secondary agent (proto-agent). For example, in the sentence $[ArgA\ John]\ walked\ [Arg0\ his\ dog]$, the predicate \textit{walk} is referring to the agent \textit{dog}, but not the primary agent \textit{John}. This scenario of proto-agent role is very rarely seen with other verbs, hence, we give lower priority to ArgA.  As per PropBank guidelines, the numbered arguments are ranked inverse to their suffix numbers.

\begin{figure}[h]
\centering
\includegraphics[width=.50\textwidth]{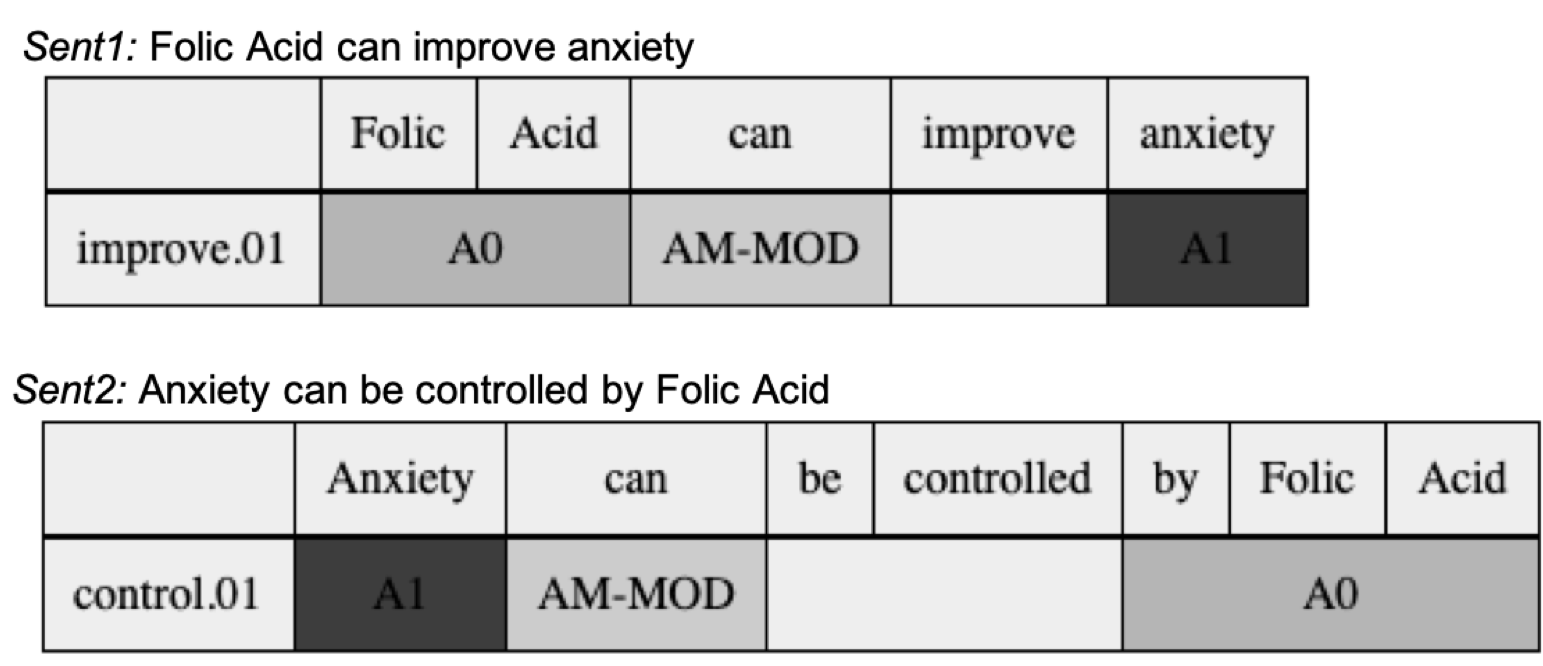}
\caption{Semantic Role Labels for the examples}

\label{fig:srlexample}
\end{figure}

There are eighteen adjunct labels, ranging from Cause and Purpose to Manner and Locative.  It is a rare phenomenon to have an adjunct as a key phrase. So a rank for adjunct is given a value lower than that for the numbered arguments. 
We show in Figure~\ref{fig:srlexample} the semantic roles with numbered arguments and adjuncts (e.g. {\it A0, A1, A2}) for the example mentioned in Section~\ref{intro}. As shown, the semantic roles of {\it Folic Acid} and {\it anxiety} are same irrespective of how the sentence is expressed. 

\section{SemanticRank}
\label{algo}

This section describes our method for key phrase extraction and ranking by using position information of key phrases in a sentence, their frequency in the document, and their semantic roles. 

\subsection{Graph Generation}
Given a document, we first do sentence tokenization and use the Stanford parser to apply part-of-speech filter, and then select, as candidate words, only the nouns and adjectives. We then use the Porter stemmer to stem all the words by removing the words' suffixes. 

Similar to previous works, a graph $G=(V,E)$ is generated for the document using the filtered words as nodes. An edge between two nodes $(v_i, v_j)$ is formed if they co-occur within a window of $k$ contiguous words in the document. Edge weight between the nodes is assigned based on the number of co-occurrences of the nodes. As the previous works have established that the direction of the edges do not impact the performance \cite{Mihalcea04TextRank}, we construct an undirected graph for simplicity.

\subsection{PageRank} 
PageRank is used in order to identify the importance of a node, based on the vertex degree \cite{page1999pagerank}. 
PageRank (PR) of a given node ($v_i$) is calculated recursively using:
\[
 PR(v_i) = \frac{(1-\lambda)}{|V|} + \lambda \sum_{v_j \to v_i} \frac{PR(v_j)}{out(v_j)}
 \]
 where, $\lambda$ is the damping factor to make sure that the process does not go into infinity loop, $|V|$ is the number of nodes in the graph, $ v_j \to v_i $ represents an edge from node $|v_j|$ to node $|v_i|$ and $out(v_j)$ represents the number of (outgoing) edges from the node $v_j$.  The $PR$ of each vertex is initialized with the node weights calculated using the $S(.)$ formulae (defined in the below section). PageRank algorithm is iterated until there is no or negligible ($\delta$) change of $PR$ scores between two successive iterations. In our case, we fixed $\lambda$ to $0.85$, $\delta$ to $0.001$ and number of iterations to 100 in the interest of time.

\subsection{Node Weight Assignment}
\label{nodeweight}

We first weigh each candidate word with its inverse sentence position in the document. The position scores are then multiplied with the frequency score to get the sentence score of a word. Mathematically, for a document, let $\phi(w)$ denote the set of candidate words, and $freq(w_i,j)$ be the frequency of $w_i \in \phi(w)$ in the $j^{th}$ sentence position. The sentence position based score $SP(.)$ of the candidate word $w_i$ is
\[
 SP(w_i) = \sum_j \frac{1}{pos_j} * freq(w_i,j) 
 \]

where, $pos_j$ is the $j$th sentence where $w_i$ appears. 

We incorporate the semantic roles of the candidate words by parsing the candidate set of sentences using SRL to get the constituents. The SRL-based score of a word $w_i$ in a sentence $s_j$ is  
 \[
 SRL(w_i,s_j) = \frac{\sum_{k=1}^{K} r_k}{K * |A|} 
 \]

where, $r_k$ is the rank of Arg $k$,  $K$ is the total number of predicates for the word $w_i$ with respect to the sentence $s_j$ and $|A|$ is the highest predicate rank that is assigned. The SRL scores of a word in each sentence is then combined to get the final SRL score of the candidate word. 
The SRL-based score $SRL(.)$ of a word $w_i$ is
\[
 SRL(w_i) = {\sum_{j=1}^{n}  SRL(w_i,s_j)} 
 \]
 where, $n$ is the number of sentences in the document where the word $w_i$ appeared.

The sentence position scores and the SRL-based scores of a word are divided by the number of sentences 
to get normalized scores. Both the scores are then multiplied to get the final relevance score of the word. The final score $S(.)$ of a node $v_i$ (candidate word)  in the graph $G = (V,E)$ is
\[
 S(v_i) = \frac{SP(w_i)}{n} * \frac{SRL(w_i)}{n} 
 \]
 where, $n$ is the number of sentences in the document where the word $w_i$ appeared.
 



 \subsection{Candidate Phrase Generation}
As in \cite{PosnRank:2017}, 
phrases in the form of regular expressions $(adjective)*(noun)+$ are considered as potential key phrases. The length of the phrases are restricted to three by allowing only unigrams, bigrams, and trigrams to increase the precision by keeping the recall constant. The overall score of a key phrase $p$ is considered as the sum of the PageRank scores of individual words $PR({w_i})$ in the phrase. The overall score of a given phrase is
\[
 Score(p) = \sum_{i=1}^{m} PR(w_i) 
 \]
where, $m$ is the total number of words in a phrase, and $PR(w_i)$ is the PageRank score of the node.


\section{Evaluation}
\label{eval}
Here, we evaluate our relevance ranking method based on its key phrase ranking efficacy on various datasets and baselines.
\subsection{Data}
The experimental datasets for key phrase ranking consisted of titles and abstracts from three datasets of scientific articles: KDD ~\cite{PosnRank:2017}, WWW~\cite{PosnRank:2017}, and Inspec~\cite{Hulth:2003}. Inspec had controlled and uncontrolled set of annotated key phrases. We used both for our analysis. 
Table \ref{tab:datasets} summarizes all the datasets, where {\it Ct\_Docs} is the number of documents in each dataset, {\it Ct\_CandPhrases} is the number of candidate phrases in each dataset, {\it Avg\_CandPhrases} is the average number of candidate phrases per document, {\it Ct\_KeyPhrases} is the number of key phrases in each dataset, {\it Avg\_KeyPhrases} is the average number of key phrases per document, {\it Ct\_Sents} is the number of sentences per document, and {\it Avg\_Sents} is the average number of sentences per document.
\begin{table}
    \centering
    \caption{Datasets used in our Experiments}
    \scriptsize
    \begin{tabular}{c|c|c|c}
         Dataset & KDD & WWW & Inspec  \\
         \hline
         Ct\_Docs & 834 & 1350 & 2000 \\
         Ct\_CandPhrases & 53642 & 80519 & 61234 \\ 
         Avg\_CandPhrases & 64.3 & 59.6 & 30.6 \\
         Ct\_KeyPhrases & 3093 & 6405 & 28222 \\
         Avg\_KeyPhrases & 3.7 & 4.7 & 14.1 \\
         Ct\_Sents & 6665 & 9896 & 13888 \\
         Avg\_Sents & 7.9 & 7.3 & 6.9 \\
         \hline
    \end{tabular}
    \label{tab:datasets}
\end{table}







\subsection{Experimental Setups}
For each experiment setup, we find precision (P), recall (R), and F1 measure (F1). Precision is the percentage of correctly extracted key phrases by the total extracted key phrases, Recall is the percentage of correctly extracted key phrases by the total author-labeled key phrases, and F1 is the harmonic mean of precision and recall. We also get the Mean Reciprocal Rank (MRR) and Mean Average Precision (MAP) scores between the ground truth key phrases and the key phrases relevance ranking ~\cite{N18-2105,Liu:2010}.



\begin{figure*}
\centering
\includegraphics[width=.30\textwidth]{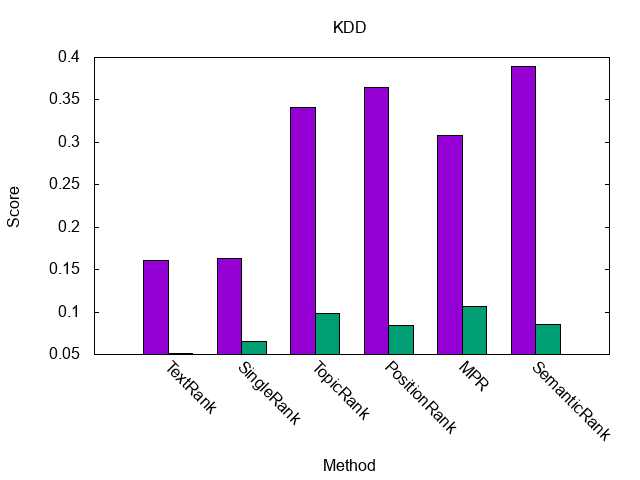}
\includegraphics[width=.30\textwidth]{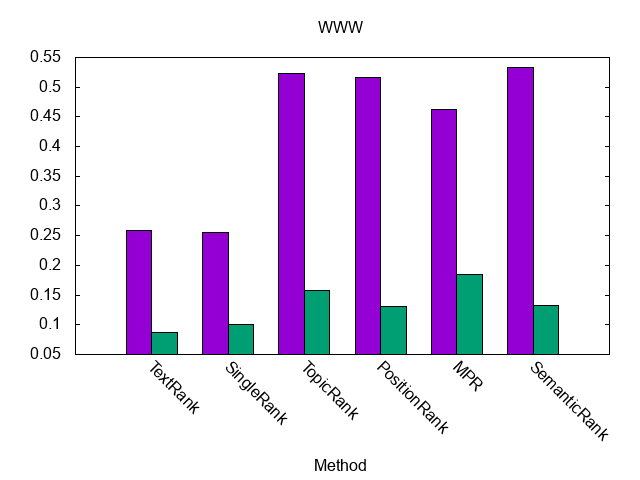}
\includegraphics[width=.32\textwidth]{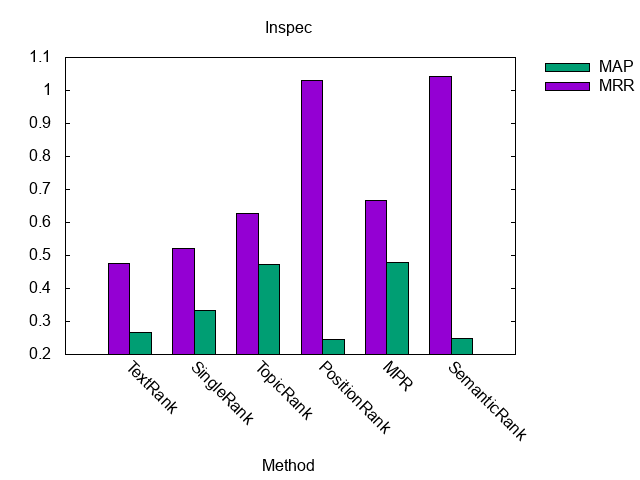}
\caption{MRR and MAP comparison for SemanticRank and baselines on the three datasets. }

\label{fig:results}
\end{figure*}

\begin{table*}[ht]
\centering
  \caption{Performance Comparison}
  \label{tab:perf2}
 \scriptsize
 \begin{tabular}{c|c|ccc|ccc|ccc|ccc|}
    \toprule
    \multicolumn{1}{c|}{Dataset} &  \multicolumn{1}{c|}{Method} & \multicolumn{3}{c|}{Top2(\%)} &  \multicolumn{3}{c|}{Top4(\%)} & \multicolumn{3}{c|}{Top6(\%)} &  \multicolumn{3}{c|}{Top8(\%)} \\
    \cline{3-14}
     & & P&R&F1&P&R&F1& P&R&F1&P&R&F1\\
    \hline
    KDD& TextRank & 8.1 & 4.0 & 5.3 & 8.3 & 8.5 & 8.1 & 8.1 & 12.3 & 9.4 & 7.6 & 15.3 & 9.8 \\
& SingleRank & 9.1 & 4.6 & 6.0 & 9.3 & 9.4 & 9.0 & 8.7 & 13.1 & 10.1 & 8.1 & 16.4 & 10.6 \\
& TopicRank & 9.3 & 4.8 & 6.2 & 9.1 & 9.3 & 8.9 & 8.8 & 13.4 & 10.3 & 8.0 & 16.2 & 10.4 \\
& PositionRank & 11.1 & 5.6 & 7.3 & 10.8 & 11.1 & 10.6 & 9.8 & 15.3 & 11.6 & 9.2 & 18.9 & 12.1 \\
& Multipartite Rank & \textbf{15.2} & \textbf{7.7} & \textbf{10.0} & 11.3 & 11.4 & 11 & 8.8 & 13.3 & 10.2 & 7.2 & 14.6 & 9.4 \\
& \textbf{ SemanticRank } & 13.1 & 6.7 & 8.6 & \textbf{ 11.5 } & \textbf{ 11.8 } & \textbf{ 11.3 } & \textbf{ 10.4 } & \textbf{ 15.9 } & \textbf{ 12.2 } & \textbf{ 9.6 } & \textbf{ 19.2 } & \textbf{ 12.8 }\\

    \hline
    WWW& TextRank & 7.7 & 3.7 & 4.8 & 8.6 & 7.9 & 8.0 & 8.1 & 12.3 & 9.8 & 8.2 & 15.2 & 10.2 \\
& SingleRank & 9.1 & 4.2 & 5.6 & 9.6 & 8.9 & 8.9 & 9.3 & 13.0 & 10.5 & 8.8 & 16.3 & 11.0 \\
& TopicRank & 8.8 & 4.2 & 5.5 & 9.6 & 8.9 & 8.9 & 9.5 & 13.2 & 10.7 & 9.0 & 16.5 & 11.2 \\
& PositionRank & 11.3 & 5.3 & 7.0 & 11.3 & 10.5 & 10.5 & 10.8 & 14.9 & 12.1 & 9.9 & 18.1 & 12.3 \\
& Multipartite Rank & \textbf{22.6} & \textbf{7.9} & \textbf{11.3} & 17 & 11.8 & 13.3 & 13.6 & 13.9 & 13.1 & 11.1 & 15 & 12.2 \\
 & \textbf{ SemanticRank } & 18.7 & 6.5 & 9.3 & \textbf{ 17 } & \textbf{ 11.8 } & \textbf{ 13.3 } & \textbf{ 15.5 } & \textbf{ 16 } & \textbf{ 15 } & \textbf{ 14.1 } & \textbf{ 19.5 } & \textbf{ 15.6 }\\

    \hline
    Inspec& TextRank & 18.7 & 3.6 & 6.0 & 16.1 & 5.3 & 8.0 & 16.3 & 5.7 & 8.5 & 17.5 & 9.5 & 12.3 \\
& SingleRank & 20.1 & 4.3 & 7.1 & 18.5 & 6.0 & 9.1 & 18.2 & 9.8 & 12.7 & 17.0 & 10.5 & 13.0 \\
& TopicRank & 25.9 & 4.4 & 7.3 & 22.6 & 7.4 & 10.7 & 20 & 9.7 & 12.5 & 18.3 & 11.7 & 13.6 \\
& PositionRank & 36.5 & 6.2 & 10.2 & 32.5 & 10.6 & 15.4 & 29.3 & 14.1 & 18.1 & 26.6 & 16.8 & 19.6 \\
& Multipartite Rank & 27.7 & 4.6 & 7.7 & 23.7 & 7.8 & 11.2 & 21 & 10.2 & 13.1 & 19 & 12.2 & 14.1 \\
 & \textbf{ SemanticRank } & \textbf{ 36.5 } & \textbf{ 6.2 } & \textbf{ 10.3 } & \textbf{ 32.5 } & \textbf{ 10.6 } & \textbf{ 15.4 } & \textbf{ 29.4 } & \textbf{ 14.1 } & \textbf{ 18.2 } & \textbf{ 26.9 } & \textbf{ 17.1 } & \textbf{ 19.8 }\\
  \bottomrule

\end{tabular}
\end{table*}




\subsection{Results}

We compare the performance of our model against five baselines: {\it TextRank}~\cite{Mihalcea04TextRank}, {\it SingleRank}~\cite{Wan:2008}, {\it TopicRank}~\cite{Liu:2010}, {\it PositionRank}~\cite{PosnRank:2017} and {\it MultipartiteRank}~\cite{N18-2105}. Table~\ref{tab:perf2} gives the performance of the different methods in terms of $P$, $R$ and $F1$ for top $N=2,4,6,8$ predicted key phrases. The best scores are highlighted in bold. On both KDD and WWW data sets, for $N=6$ and $N=8$, our method has an improvement in F1 score over all baseline methods.  For the WWW dataset, there is a statistically significant improvement in F1 score over all the baseline methods ($p \leq 0.01$). For the KDD dataset, a p-value of $0.09$ and $0.07$ was obtained for $N=6$ and $N=8$ respectively, indicating that semantic roles influence the ranking of key phrases but significantly only at $10\%$ level of significance. For $N=4$, our method performs on-par with {\it MultipartiteRank} on both KDD and WWW. Finally, on the Inspec dataset, our method achieves competitive and comparable results with {\it PositionRank}. 

Figure~\ref{fig:results} shows, in a bar chart form, the MRR and MAP values of all the methods on the three datasets. As shown in the figure, for both KDD and WWW datasets, {\it SemanticRank} acheived better MRR score than the other baseline methods. For Inspec, {\it SemanticRank}'s performance is comparable to PositionRank.  The MAP scores for {\it TopicRank} and {\it Multipartite Rank (MPR)} are high, as they filter out and present only representative key phrases for each topic of candidate phrases,  where as other techniques try to assign scores for each candidate key phrase.

We assigned different ranks to different roles in order of their importance, with `Arg0' having the maximum rank, `Arg1' having the next maximum rank, and had assigned ranks to all other roles in decreasing order of ranks. We changed the ranks of these roles to test the sensitivity of the method. We assigned same ranks to roles `Arg0' and `Arg1'; for the other roles we kept the ranks as same as before ({\it SentSRL1}). We also experimented with a setting where 'Arg0' and `Arg1' roles had the same high rank, `Arg2', `Arg3' and `Arg4' have the same low rank and rest of the roles are assigned a very small rank ({\it SentSRL2}).  Table ~\ref{tab:SAperf2} shows the precision, recall and F1 measure performance comparison of the different combinations of semantic roles on the KDD, WWW and Storwize datasets. For the different combinations of rank assignment to the semantic roles, the precision, recall, and F1 measures are practically the same for KDD and WWW. This indicates that the key phrase ranking method is rather insensitive to variation in the semantic roles' rank values.

\section{Conclusions and Future Work}
\label{conclu}
This paper introduced a linguistically motivated approach to key phrase relevance ranking based on semantic roles of words. 
The method considered the candidate phrase sentence position and its semantic roles for relevance ranking of phrases. Experimental results show that using semantic role knowledge can effectively improve the quality of key phrases extracted from documents. 

Limitations of our approach include: (1) The SRL implementation that we used for our experiments has a precision of 87\% for in-domain documents and 77\% for out-of-domain documents.
(2) SRL works well for well-formed and grammatically correct sentences, and gives inferior results for texts that are not grammatically correct. In scientific articles title holds key information, but are not grammatically complete/correct most of the time.
In spite of the limitations in the realization of SRL, sentence-based position information along with SRL gave improved results for well structured datasets.  
In future work we plan to investigate approaches to integrate semantic roles with the other features, possibly using neural networks to combine them.
\newpage

\bibliographystyle{acl_natbib}
\bibliography{acl2019}

\begin{thebibliography}{13}
\expandafter\ifx\csname natexlab\endcsname\relax\def\natexlab#1{#1}\fi

\bibitem[{Boudin(2018)}]{N18-2105}
Florian Boudin. 2018.
\newblock \href {https://doi.org/10.18653/v1/N18-2105} {Unsupervised keyphrase
  extraction with multipartite graphs}.
\newblock In \emph{Proceedings of the 2018 Conference of the North American
  Chapter of the Association for Computational Linguistics: Human Language
  Technologies, Volume 2 (Short Papers)}, pages 667--672. Association for
  Computational Linguistics.

\bibitem[{Bougouin et~al.(2013)Bougouin, Boudin, and Daille}]{graph2}
A.~Bougouin, F.~Boudin, and B.~Daille. 2013.
\newblock Topicrank: Graph-based topic ranking for keyphrase extraction.
\newblock In \emph{International Joint Conference on Natural Language
  Processing}, pages 543--551.

\bibitem[{Danesh et~al.(2015)Danesh, Sumner, and Martin}]{graph3}
S.~Danesh, T.~Sumner, and J.~H. Martin. 2015.
\newblock Sgrank: Combining statistical and graphical methods to improve the
  state of the art in unsupervised keyphrase extraction.
\newblock \emph{Lexical and Computational Semantics}.

\bibitem[{Florescu and Caragea(2017)}]{PosnRank:2017}
Corina Florescu and Cornelia Caragea. 2017.
\newblock \href {https://doi.org/10.18653/v1/P17-1102} {Positionrank: An
  unsupervised approach to keyphrase extraction from scholarly documents}.
\newblock In \emph{Proceedings of the 55th Annual Meeting of the Association
  for Computational Linguistics (Volume 1: Long Papers)}, pages 1105--1115.
  Association for Computational Linguistics.

\bibitem[{Gildea and Jurafsky(2002)}]{Gildea}
Daniel Gildea and Daniel Jurafsky. 2002.
\newblock \href {https://doi.org/10.1162/089120102760275983} {Automatic
  labeling of semantic roles}.
\newblock \emph{Comput. Linguist.}, 28(3):245--288.

\bibitem[{Hasan and Ng(2014)}]{hasan-ng:2014}
Kazi~Saidul Hasan and Vincent Ng. 2014.
\newblock \href {http://www.aclweb.org/anthology/P14-1119} {Automatic keyphrase
  extraction: A survey of the state of the art}.
\newblock In \emph{Proceedings of the 52nd Annual Meeting of the Association
  for Computational Linguistics (Volume 1: Long Papers)}, pages 1262--1273,
  Baltimore, Maryland. Association for Computational Linguistics.

\bibitem[{Hulth(2003)}]{Hulth:2003}
Anette Hulth. 2003.
\newblock \href {https://doi.org/10.3115/1119355.1119383} {Improved automatic
  keyword extraction given more linguistic knowledge}.
\newblock In \emph{Proceedings of the 2003 Conference on Empirical Methods in
  Natural Language Processing}, EMNLP '03, pages 216--223, Stroudsburg, PA,
  USA. Association for Computational Linguistics.

\bibitem[{Liu et~al.(2010)Liu, Huang, Zheng, and Sun}]{Liu:2010}
Zhiyuan Liu, Wenyi Huang, Yabin Zheng, and Maosong Sun. 2010.
\newblock \href {http://dl.acm.org/citation.cfm?id=1870658.1870694} {Automatic
  keyphrase extraction via topic decomposition}.
\newblock In \emph{Proceedings of the 2010 Conference on Empirical Methods in
  Natural Language Processing}, EMNLP '10, pages 366--376, Stroudsburg, PA,
  USA. Association for Computational Linguistics.

\bibitem[{Mihalcea and Tarau(2004)}]{Mihalcea04TextRank}
R.~Mihalcea and P.~Tarau. 2004.
\newblock {TextRank}: Bringing order into texts.
\newblock In \emph{Proceedings of {EMNLP-04}and the 2004 Conference on
  Empirical Methods in Natural Language Processing}.

\bibitem[{Page et~al.(1999)Page, Brin, Motwani, and
  Winograd}]{page1999pagerank}
Lawrence Page, Sergey Brin, Rajeev Motwani, and Terry Winograd. 1999.
\newblock The pagerank citation ranking: Bringing order to the web.
\newblock Technical report, Stanford InfoLab.

\bibitem[{Pradhan et~al.(2004)Pradhan, Ward, Hacioglu, Martin, and
  Jurafsky}]{pradhan2004shallow}
Sameer~S Pradhan, Wayne~H Ward, Kadri Hacioglu, James~H Martin, and Daniel
  Jurafsky. 2004.
\newblock Shallow semantic parsing using support vector machines.
\newblock In \emph{HLT-NAACL}, pages 233--240.

\bibitem[{Roth and Woodsend(2014)}]{roth2014composition}
Michael Roth and Kristian Woodsend. 2014.
\newblock Composition of word representations improves semantic role labelling.
\newblock In \emph{Proceedings of the 2014 Conference on Empirical Methods in
  Natural Language Processing (EMNLP)}, pages 407--413.

\bibitem[{Wan and Xiao(2008)}]{Wan:2008}
Xiaojun Wan and Jianguo Xiao. 2008.
\newblock Single document keyphrase extraction using neighborhood knowledge.
\newblock In \emph{Proceedings of the 23rd National Conference on Artificial
  Intelligence - Volume 2}.

\end{thebibliography}

\end{document}